\documentclass[osajnl,twocolumn,showpacs,superscriptaddress,10pt]{revtex4-1}
\usepackage{amsmath,amssymb,graphicx}
\begin{document}

\title{Active plasmon injection scheme for subdiffraction imaging with imperfect negative index flat lens}

\author{Anindya Ghoshroy}
\author{Wyatt Adams}
\author{Xu Zhang}
\author{Durdu \"O. G\"uney}\email{Corresponding author: dguney@mtu.edu}
\affiliation{Department of Electrical and Computer Engineering , Michigan Technological University, 1400 Townsend Dr, Houghton, MI 49931-1295, USA}

\begin{abstract}
We present an active physical implementation of the recently introduced plasmon injection loss compensation scheme for Pendry's non-ideal negative index flat lens in the presence of realistic material losses and signal-dependent noise. In this active implementation, we propose to use a physically convolved external auxiliary source for signal amplification and suppression of the noise in the imaging system. In comparison with the previous passive implementations of the plasmon injection scheme for sub-diffraction limited imaging, where an inverse filter post-processing is used, the active implementation proposed here allows for deeper subwavelength imaging far beyond the passive post-processing scheme by extending the loss compensation to even higher spatial frequencies.
\end{abstract}

\maketitle

\section{Introduction}

Controlling the interaction of photons and electrons at the sub-wavelength electromagnetic regime has led to a wide variety of novel optical materials
and applications in the territory of metamaterials and plasmonics relevant to computing, communications, defense, health, sensing, imaging, energy,
and other technologies \cite{Fang534,taubner2006near,zhang2008superlenses,liu2007far,rho2010spherical,lu2012hyperlenses,sun2015experimental,lee2007development,valentine2008three,landy2008perfect,temnov2012ultrafast,aslam2012negative,sadatgol2016enhanced,choi2011terahertz,chen2012extremely,zhang2015extremely,schurig2006metamaterial,gwamuri2013advances,rockstuhl2008absorption,vora2014exchanging,vora2014multi,bulu2005compact,odabasi2013electrically,guney2009negative,smolyaninov2010metric,tame2013quantum,al2015quantum,asano2015distillation,jha2015metasurface,sperling2008biological,huang2006cancer,lal2008nanoshell,guo2010multifunctional,doi:10.1021/nl304208s,blankschien2012light,chen2010enhanced,sherlock2011photothermally,ahmadivand2015enhancement}.
The prospect of circumventing Rayleigh's diffraction limit, thereby allowing super-resolution imaging has regained tremendous ground since Pendry
theorised that a slab of negative (refractive) index material (NIM) can amplify and focus evanescent fields which contain information about the
sub-wavelength features of an object \cite{Pendry}. A recent review of super-resolution imaging in the context of metamaterials is given in \cite{adams2016review}.

However, a perfect NIM does not exist in nature and although recent developments in metamaterials have empowered their realization, a fundamental
limitation exists. The presence of material losses in the near infrared and visible region is significant \cite{PhysRevLett.95.137404,RN4,Dolling892}.
This compromises the performance of the theoretical perfect lens \cite{Webb1,Yang:05} since a significant portion of evanescent fields is below the
noise floor of the detector and are indiscernible. Therefore, new efforts were directed towards the compensation of losses in metamaterials \cite{Nezhad:04,Popov:06,PhysRevB.80.125129}.
Amongst the schemes that were developed, gain media to compensate intrinsic losses  gained popularity \cite{Anantha,refId0}. However,
Stockman \cite{PhysRevLett.98.177404} demonstrated that the use of gain media involved a fundamental limitation. Using the Kramers-Kronig relations,
they developed a rule based on causality which makes loss compensation with gain media difficult to realize.

Recently, a new compensation scheme, called plasmon injection or $\Pi$ scheme \cite{PI}, was proposed. The $\Pi$ scheme was conceptualized with
surface plasmon driven NIMs \cite{Aslam} and achieves loss compensation by coherently superimposing externally injected surface plasmon polaritons (SPPs) with local SPPs.
Therefore, absorption losses in the NIM could be removed without a gain medium or non-linear effects. Although the $\Pi$  scheme was originally
envisioned for plasmonic metamaterials \cite{PI,guney2011surface,aslam2012dual}, the idea is general and can be applied to any type of optical modes.
In \cite{Wyatt} Adams, et. al used a post processing technique equivalent to this method.  They demonstrated that the process can indeed be used to
amplify the attenuated Fourier components and thereby accurately resolve an object with sub-wavelength features.

\begin{figure}[htbp]
\centering
\includegraphics[width=\linewidth,height=5cm]{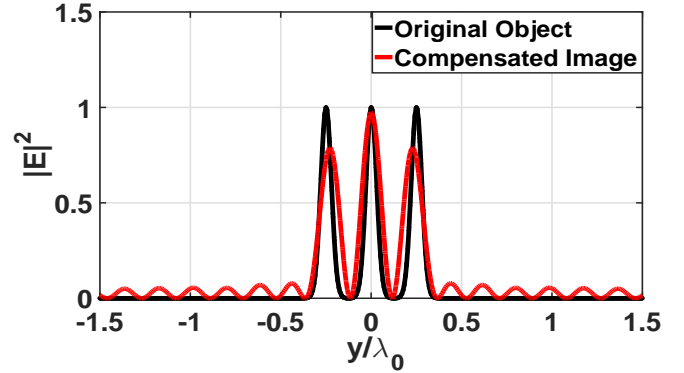}
\caption{Electric field magnitude squared $[Vm^{-1}]^2$ distribution in the object and image planes for an object with features separated by $\lambda _o/4$ with $\lambda _o = 1 \mu m$. The compensated image is reasonably well resolved with the equivalent inverse filter post-processing technique. }
\label{fig:Fig1}
\end{figure}

\begin{figure}[htbp]
\centering
\includegraphics[width=\linewidth]{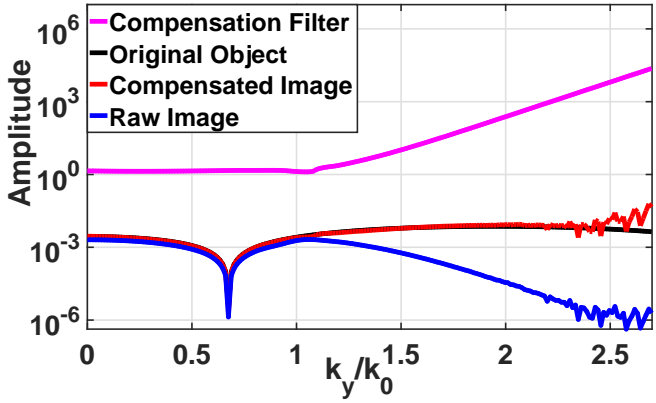}
\caption{Fourier spectra $[Vm^{-1}]$ of the three Gaussians and the
compensated image in figure \ref{fig:Fig1}, and the raw image obtained
without loss compensation. The compensation filter is the inverse
of the transfer function. The compensated image spectrum is
obtained simply by multiplying the raw image spectrum with the
compensation filter. Notice that the noise can be seen for high
spatial frequencies which is amplified by the compensation.}
\label{fig:Fig2}
\end{figure}
\begin{figure}[htbp]
\centering
\includegraphics[width=\linewidth,height=5cm]{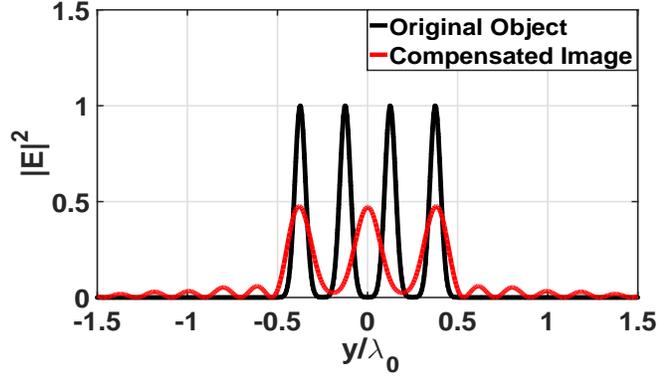}
\caption{Electric field magnitude squared $[Vm^{-1}]^2$ for the four Gaussians, in the object and image planes, separated by $\lambda _o/4$ with $\lambda _o = 1 \mu m$. The compensated image  is very poorly resolved with one of the Gaussians missing.}
\label{fig:Fig3}
\end{figure}

\begin{figure}[htbp]
\centering
\includegraphics[width=\linewidth]{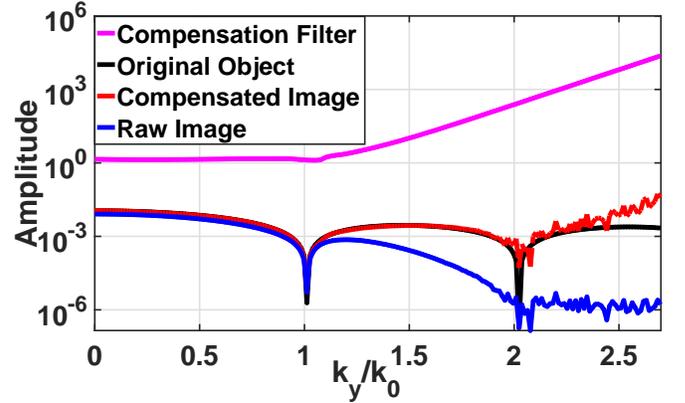}
\caption{Fourier spectra $[Vm^{-1}]$ of the four Gaussians and the compensated image in figure \ref{fig:Fig3}, and the raw image along with the
compensation filter. Notice that the feature at $\frac{k_y}{k_o} = 2$ is not discernable under the noise and cannot be recovered well with the compensation.}
\label{fig:Fig4}
\end{figure}

Although this form of passive inverse filter provides compensation for absorption losses, it is also prone to noise amplification \cite{Wyatt}. This
is illustrated in figure \ref{fig:Fig1} which shows an object with three Gaussian features separated by $\lambda _o/4$, where $\lambda _o$ is the
free space wavelength. Noise is prominent in the Fourier spectra beyond $\frac{k_y}{k_o} = 2.5$ as seen in figure \ref{fig:Fig2}. However, the
compensated image is still reasonably well resolved. Consider now the object shown in figure \ref{fig:Fig3}, which has four Gaussians separated
by $\lambda _o/4$. The Fourier spectra of the raw image, shown in figure \ref{fig:Fig4}, demonstrates how the feature at $\frac{k_y}{k_o} = 2$ is
not distinguishable under the noise. The final compensated image, when subject to the same compensation scheme, is poorly resolved. We define
a \textbf{\emph{feature}} as any spatial Fourier component that has substantial contribution to the shape of the object. It is clear that the
Fourier components beyond $\frac{k_y}{k_o} = 2$ have a significant contribution to the four Gaussians and must be recovered from the image spectrum
in order to accurately resolve the object. Therefore, noise presents a limitation which must be overcome to make the $\Pi$ scheme versatile.

In the present work, we demonstrate how the $\Pi$ scheme can be significantly improved with the use of a physical auxiliary source to recover
high spatial frequency features that are buried under the noise. We show that by using a convolved auxiliary source we can amplify the object
spectrum in the frequency domain. The amplification makes the Fourier components that are buried in the noise distinguishable. This allows for
the recovery of the previously inaccessible object features by adjusting the amount of compensation from the $\Pi$ scheme.

The technique presented in this paper is based on the same negative index flat lens (NIFL) as in \cite{Wyatt}. We use the words
\textbf{\emph{"passive"}} and \textbf{\emph{"active"}} to distinguish between the compensation schemes applied in \cite{Wyatt} and in this work,
respectively. Therefore, the inverse filter post processing used in \cite{Wyatt} and figures \ref{fig:Fig1}-\ref{fig:Fig4} to emulate the physical
compensation of losses can be called passive $\Pi$ scheme, since no external physical auxiliary is actively involved as opposed to the active
$\Pi$ scheme here, where the direct physical implementation using an external auxiliary source as originally envisioned in \cite{PI} is sought.
The active compensation scheme allows us to control noise amplification and hence extend the applicability of the $\Pi$ scheme to higher spatial
frequencies.

\section{Theory}

\begin{figure}[htbp]
\centering
\includegraphics[width=\linewidth]{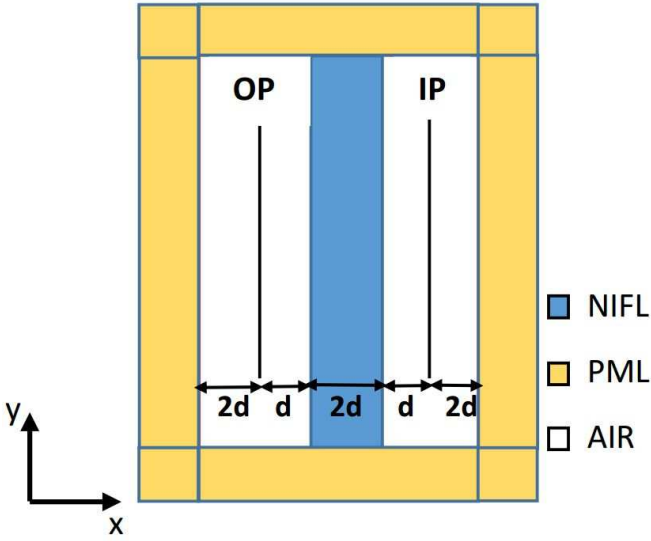}
\caption{The geometry built in COMSOL to perform numerical simulations (not to scale). OP and IP are the object and image planes, respectively. Electric field is polarized along the z-axis (pointing out of plane). The object is defined as an electric field distribution $[E_z(y)]$ on the object plane. The operating wavelength is $\lambda _o = 1 \mu m$ and $2d = 0.5 \mu m$.  Blue, white,
and orange regions are the NIFL, air, and perfectly matched layer (PML), respectively. }
\label{fig:Fig5}
\end{figure}

We define the optical properties of the NIFL with the relative permittivity and permeability expressed
as $\epsilon_r = \epsilon^{'} + i\epsilon^{''}$ and $\mu_r = \mu^{'} + i\mu^{''}$, where $\epsilon^{'} = -1$ and $\mu^{'} = -1$.
COMSOL Multiphysics, the finite element method based software package that we use here, assumes $exp(j\omega t)$ time dependence.
Therefore, the imaginary parts of $\epsilon_r$ and  $\mu_r$ are negative for passive media. In this paper we have used $0.1$ as the imaginary
parts of both $\epsilon_r$ and  $\mu_r$ which is a reasonable value given currently fabricated metamaterial structures \cite{RN1,Garc,Verhagen}.
The geometry used to numerically simulate the NIFL in COMSOL is given in figure \ref{fig:Fig5}. The first step is characterizing the NIFL with
a transfer function. For a detailed discussion on the geometry setup and transfer function  calculations, the reader is referred to \cite{Wyatt}.
Here, we present a brief mathematical description of the compensation scheme.

The spatial Fourier transforms of the electric fields in the object and image planes are related by the passive transfer function,
$T_P(k_y)$ of the imaging system, which can be calculated with COMSOL. This is expressed mathematically as
\begin{equation}
I(k_y) = T_P(k_y)O(k_y).
\label{eq:TF}
\end{equation}
Here $O(k_y) = \mathcal{F}\{O(y)\}$ and $I(k_y) = \mathcal{F}\{I(y)\}$, where $O(y)$ and $I(y)$ are the spatial distribution
of the electric fields in the object and image planes, respectively, and $\mathcal{F}$ is the Fourier transform operator. According
to  \cite{Wyatt} the passive compensation is defined by the inverse of the transfer function. Hence, the loss compensation is achieved
by multiplying the raw image spectrum in Eq. \ref{eq:TF} with the inverse of the transfer function given by
\begin{equation}
C_P(k_y) = \bigg[T_P(k_y)\bigg]^{-1}.
\label{eq:CF}
\end{equation}

For \emph{"active"} compensation we first define a mathematical expression given by
\begin{equation}
A(k_y) = 1+ P(k_y),
\label{eq:AUX}
\end{equation}
\begin{equation}
P(k_y) = P_oexp\bigg[- \frac{(\frac{k_y}{k_o} - k_c)^2}{2\sigma ^2}\bigg],
\label{eq:PUMP}
\end{equation}
where $P_o$ is a constant. $k_c$ controls the center frequency of the Gaussian, $k_o = \frac{2\pi}{\lambda}$ is the free space wave number
and $\sigma$ controls the full width at half maximum (FWHM) of the Gaussian. We convolve $A(y) = \mathcal{F}^{-1}\{A(k_y)\}$ with the object $O(y)$ in the spatial domain and denote the new object by $O^{'}(y)$. This is expressed as
\begin{equation}
O^{'}(y) = \int\limits^{\infty}_{-\infty} O(y)A(y-\alpha) d\alpha .
\label{eq:CONV}
\end{equation}
We shall refer to this convolved object as the \emph{total object}. Since convolution in the spatial domain is equivalent to multiplication
in the spatial frequency domain, the Fourier spectrum of the total object $O^{'}(k_y) = \mathcal{F}\{O^{'}(y)\}$, is related to the original
object by
\begin{equation}
O^{'}(k_y) =  O(k_y) + O(k_y)P(k_y).
\label{eq:TOTALOBJ}
\end{equation}
The second term on the RHS will be referred to as the \emph{"auxiliary source,"} where $P_o$ in Eq. \ref{eq:PUMP} defines its amplitude at the
center frequency  $k_c$ . Note that this term, which is a convolution of $O(y)$ with $P(y)= \mathcal{F}^{-1}\{P(k_y)\}$, represents amplification
in the spatial frequency domain provided that $P(k_y) > 1$. Even though the auxiliary source is object dependent, as we will discuss later,
the external field to generate auxiliary source does not require prior knowledge about the object. Now, the Fourier transform of the fields
in the object and image planes are related to each other by the transfer function of the NIFL as defined by Eq. \ref{eq:TF}. Therefore, in
response to the total object, the new field distribution in the image plane, expressed as $I^{'}(y) = \mathcal{F}^{-1}\{I^{'}(k_y)\}$, is
transformed as
\begin{equation}
I^{'}(k_y) =  T_P(k_y)O^{'}(k_y),
\label{eq:IMGTOTALOBJ}
\end{equation}
where we can plug in the value of $O^{'}(k_y)$ from Eq. \ref{eq:TOTALOBJ} to obtain the convolved image
\begin{equation}
I^{'}(k_y) =  T_P(k_y)O(k_y) + T_P(k_y)O(k_y)P(k_y).
\label{eq:IMGTOTALOBJ1}
\end{equation}
\begin{figure}[h]
\centering
\includegraphics[width=\linewidth,height=5cm]{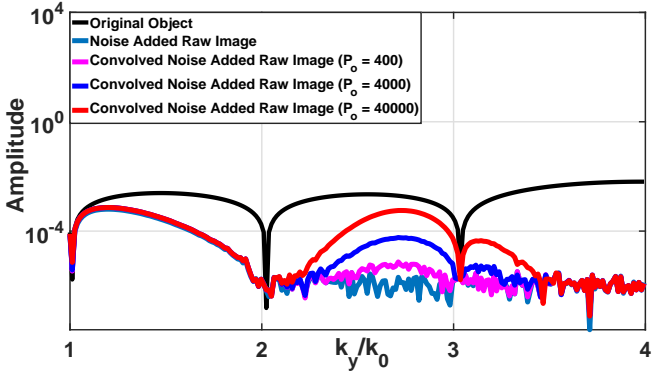}
\caption{Fourier spectra [$Vm^{-1}$] of an arbitrary object illustrating  how the auxiliary amplitude can be tuned from the object plane to raise the image spectrum above the noise floor by controlling the amplification. $P(k_y)$ is centered at $k_c = \frac{k_y}{k_o} = 3$ with $\sigma = 0.13.$}
\label{fig:Fig6}
\end{figure}

\noindent The second term on the RHS of Eq. \ref{eq:IMGTOTALOBJ1} is a measure of the residual amplification  which managed to propagate to the image plane.
Therefore, by controlling $P_o$, from the object plane, we can tune the necessary amplification of high spatial frequency features to raise the
desired frequency spectrum above the noise floor in the image plane. This process is illustrated in figure \ref{fig:Fig6} for different auxiliary
amplitudes.

The new "active" loss compensation  scheme must consider the extra power that is now available in the image spectrum. We distinguish the compensation
scheme from Eq. \ref{eq:CF} with the subscript "A". We start by defining the active transfer function of the NIFL as
\begin{equation}
T_{A}(k_y) = \frac{I^{'}(k_y) }{O(k_y) }.
\label{eq:ACT}
\end{equation}
The numerator of Eq. \ref{eq:ACT} is the image of the total object which is given by Eq. \ref{eq:IMGTOTALOBJ1}. This transfer function is called
"active" because it considers the auxiliary to be a part of the imaging system. Plugging in the value of $I^{'}(k_y)$ from Eq. \ref{eq:IMGTOTALOBJ1}
into  Eq. \ref{eq:ACT} we obtain the following expression for the active transfer function,
\begin{equation}
T_{A}(k_y) = T_{P}(k_y)+  T_{P}(k_y)P(k_y).
\label{eq:TFA}
\end{equation}
The active compensation filter is simply defined as the inverse of the active transfer function and is expressed mathematically by
\begin{equation}
C_{A}(k_y) = \bigg[T_{P}(k_y)+  T_{P}(k_y)P(k_y)\bigg]^{-1}.
\label{eq:CFA}
\end{equation}

\begin{figure}[h]
\centering
\includegraphics[width=\linewidth,height=5cm]{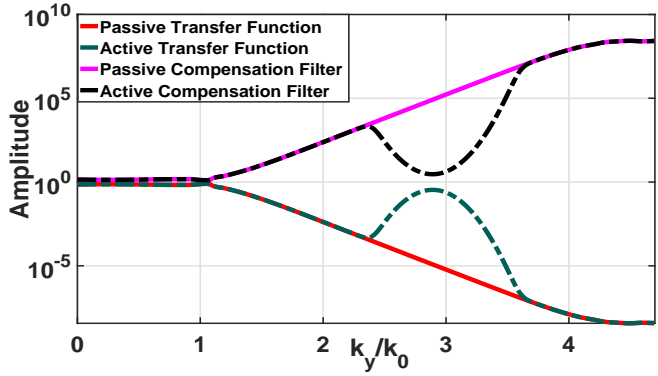}
\caption{Comparisons of the passive transfer function $T_P(k_y)$ and compensation filter $C_P(k_y)$ with the active transfer function $T_A(k_y)$ and compensation filter $C_A(k_y)$. $P(k_y)$ incorporated into the active compensation filter is centered at  $k_c = \frac{k_y}{k_o}= 3$ with $ \sigma = 0.13$.}
\label{fig:Fig7}
\end{figure}
Figure \ref{fig:Fig7} illustrates the active and passive transfer functions and the corresponding loss compensation schemes. The amount of active
compensation drops within $2.5 < \frac{k_y}{k_o} < 3$. This indicates that in this region the auxiliary source is expected to provide compensation
to the image. Therefore, the greater the auxiliary power, the lower is the required compensation through inverse filter within that region of spatial
frequencies. It is interesting to note at this point the similarity of the active transfer functions in figure \ref{fig:Fig7} and those in
\cite{Chen:16}. In the latter, however, highly stringent conditions are imposed on the negative index lens to obtain such a transfer function.

\section{Noise Characterization}

The active compensation scheme will be applied to an NIFL imaging system affected by noise where the noise process is a circular Gaussian random
variable. Although there are many different sources of noise, they can be  broadly classified into, "signal-dependent" (SD) and
"signal-independent" (SI). The random nature of noise manifests itself in the form of an uncertainty in the level of the desired signal. This
uncertainty is quantified by the standard deviation $\sigma _n$.  The actual distortion can be thought of as a random selection from an infinite
set of values and the selection process obeys a probability distribution function. The standard deviation describes the range of values which have
the greatest likelihood of being selected. When the underlying signal is distorted by multiple independent sources of noise, each characterized by
Gaussian distributions, then the variance $(\sigma _n ^{2})$ of the total noise is the sum of the variances of individual noise sources \cite{Fiete}.

SD noise, as the name implies, is characterized by a $\sigma _n$ that is intricately related to spatial (or temporal) variations in the incoming
signal intensity. The magnitude of the signal distortion therefore also increases with the signal strength. Sources of SD noise in an imaging system
can be present on the detector side or the transmission medium. For example, the statistical nature of photons manifests itself as noise which has
Poissonian statistics. In radiographic detection equipment, such sources of noise are called quantum mottle or quantum
noise \cite{0031-9155-48-23-006,MP:MP5126}. Another source of SD noise originates from roughness of the transmission media, which in sub-wavelength
imaging systems can be for example, surface roughness of the NIFL. One can think of surface irregularities as electromagnetic scatterers which radiate in different directions, distorting the propagating wave. Previous experiments on the impact of surface roughness \cite{Guo2014,Wang:11,Liu_Hong} showed that increasing material losses in the NIFL improved the image resolution of the perfect lens for relatively large surface roughness. Although this may seem counter-intuitive, it can be explained if roughness is modelled as a source of scattering. Adding material loss is equivalent to lowering the power transmission of the lens. This lowers the magnitude of the excitation field responsible for scattering effects and in turn reduces the magnitude of the scattered field. If material losses are kept constant, the scattering process will be proportional to the intensity of illumination provided to the object. Therefore, such kind of noise is amplified as the illumination intensity is increased. On the other hand, the SI noise is quantified by a standard deviation which is not a function of the incoming signal. Therefore, the random nature of the noise will be visible only when the incoming signal amplitude is comparable to the distortions due to the SI noise. A good example of this is "dark noise", which affects a CCD sensor even in the absence of illumination \cite{holst1998ccd}.

A well known model \cite{Walkup} used to describe the spatial distribution of a signal that has been distorted with both SD and SI sources of noise
is
\begin{equation}
r(y) = s(y) + f(s(y))N_{1}(y) + N_{2}(y),
\label{eq:NOISE}
\end{equation}
where $s(y)$ is the noiseless or ideal signal and $r(y)$ is the noisy version. $N_{1}(y)$ and $N_{2}(y)$ are two statistically independent random
noise processes with zero mean and Gaussian probabilities with standard deviations $\sigma _{n1}$ and $\sigma _{n2}$, respectively. The noise
processes $N_{1}(y)$ and $N_{2}(y)$ are signal independent. The signal dependent nature of noise is modelled by modulating $N_{1}(y)$  using the
function $f(s(y))$. Generally, $f(s(y))$ is a non-linear function of the ideal signal itself which is chosen based on the system which
Eq. \ref{eq:NOISE} is attempting to describe. For example, $f(s(y))$ is usually considered to be the photographic density which is unitless when
modelling signal-dependent film grain noise. Therefore, $f(s(y))N_{1}(y)$ represents the effective signal dependent noise term. We can re-write the
expression in Eq. \ref{eq:NOISE} as
\begin{equation}
r(y) = s(y) + N_{SD}(y) + N_{SI}(y),
\label{eq:NOISE1}
\end{equation}
where the standard deviation of $N_{SD}(y)$ is $f(s(y))\sigma _{n1}$ and the subscripts SD, SI distinguish between the sources of noise. The noise
model of Eq. \ref{eq:NOISE}, referred to as the \emph{signal modulated noise model}, is used for signal estimation purposes with the Wiener filter.
A detailed discussion on this can be found in \cite{Heine:06,Kasturi:83,Froehlich:81,Walkup}. However, we will use Eq. \ref{eq:NOISE1} in this paper
for mathematical convenience to analyse the relative contributions of SD and SI noise.

In the NIFL imaging system which we consider, the ideal signal $s(y)$ will be the electric field distribution on the image plane, that is $I(y)$
and $I^{'}(y)$ for passive and active schemes, respectively. In \cite{Chen:16}, Chen, et. al adopted a $60 dB$ signal to noise ratio (SNR) in their
negative index lens considering an experimental imaging system detector \cite{Akiba:10}. This corresponds to a SD standard deviation of $10^{-3}I(y)$. In this work, we adopt the same standard for the SD noise. Additionally, we assume a SI noise process in the imaging system by adopting a spatially invariant standard deviation of $10^{-3} [V/m]$. This means that even in the absence of illumination, there is a constant background noise of the order of $1 \ mV/m$ in the detector. Although the value of the SI noise is chosen arbitrarily, this does not limit the results discussed in this paper, since the SI noise can be easily suppressed by additional auxiliary power.

By taking into consideration the SNR standard used by Chen, et. al \cite{Chen:16} and the mathematical form of Eq. \ref{eq:NOISE1} we can frame the
equations for the noisy images as
\begin{equation}
I_N(y) = I(y) + N_{SD}(y) + N_{SI}(y)
\label{eq:NOISE_PASSIVE}
\end{equation}
and
\begin{equation}
I^{'}_N(y) = I^{'}(y) + N^{'}_{SD}(y) + N^{'}_{SI}(y)
\label{eq:NOISE_ACTIVE}
\end{equation}
corresponding to the ideal images described by Eqs. \ref{eq:TF} and \ref{eq:IMGTOTALOBJ1}, respectively.  The subscript N indicates the noisy image.
The standard deviations of the noise processes are
\begin{equation}
\sigma _{n(SD)} = 10^{-3}I(y),
\label{eq:SD_Passive}
\end{equation}
\begin{equation}
\sigma ^{'} _{n(SD)} =  10^{-3}I^{'}(y)
\label{eq:SD_Active}
\end{equation}
and
\begin{equation}
\sigma _{n(SI)} = \sigma^{'} _{n(SI)} = 10^{-3} \ V/m.
\label{eq:SD_SI}
\end{equation}
Eqs. \ref{eq:SD_Passive} - \ref{eq:SD_SI} fully describe the random variables that are used to construct the SI and SD noise terms in
Eqs. \ref{eq:NOISE_PASSIVE} and \ref{eq:NOISE_ACTIVE}.

\begin{figure}[hpbp]
\centering
\includegraphics[width=\linewidth]{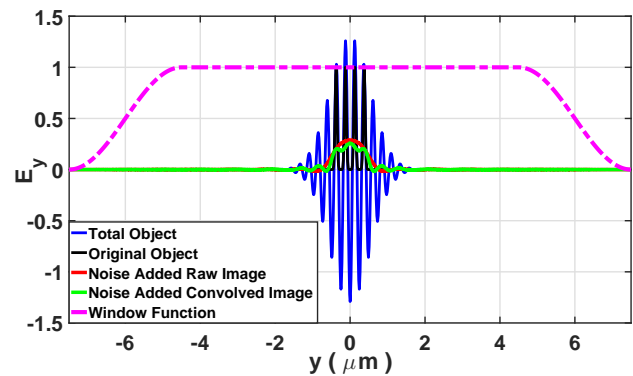}
\caption{Electric field [$Vm^{-1}$] distributions in the object and image planes. The fields on the image planes are multiplied by the window function to reduce the errors in the Fourier transform. The total object has been scaled down by $\approx 10^4$.}
\label{fig:Fig8}
\end{figure}

\section{Results}

Having described how both SD and SI noise are added to the system, the next step is to evaluate the performance of active-compensation and compare
with the passive version. We will attempt to image the previously exemplified object comprising four Gaussian features, separated
by $\frac{\lambda _o}{4}$ with $\lambda _o = 1 \mu m$ and compare the results of passive and active compensation. Note that due to the finite
extent of the image plane, it is necessary to multiply the electric fields with a window function to ensure that the field drops to zero where
the image plane is abruptly terminated. Otherwise, errors are introduced in the Fourier transform calculations. Windowing the field distribution
simply reduces these sources of error, which will be then visible only in the higher spatial frequencies. Since these errors are very small
compared with the amplitude of the SD and SI sources of noise, they do not have a significant impact on the calculations. Increasing the length
of the image plane along the y-axis can also reduce these errors, but because of computational constraints this may not be desirable.

\begin{figure}[htbp]
\centering
\includegraphics[width=\linewidth]{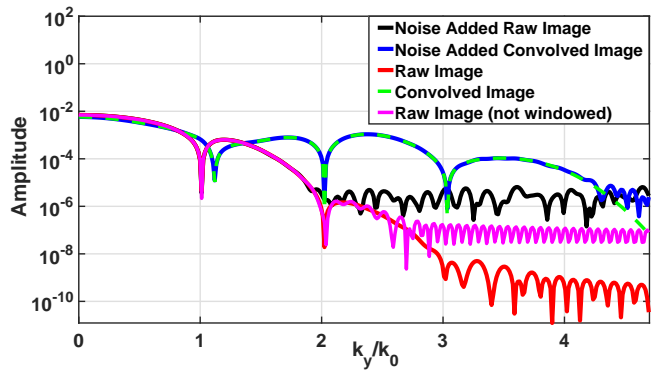}
\caption{Amplitude of the Fourier transforms [$Vm^{-1}$] on a log scale. Red and green lines are the image spectra $I(k_y)$ and $I^{'} (k_y)$ with no added noise, respectively. The blue line shows the image of the total object with added noise, $I^{'}_{N}(k_y)$. Noise is visible at $\frac{k_y}{k_o} \ > \ 4.5$ due to the inadequate amplification. The apparent noise in the pink line are due to the numerical errors introduced by the Fourier transform and shifts to higher $ \frac{k_y}{k_o}$ values after applying the Tukey window as seen in the red line.}
\label{fig:Fig9}
\end{figure}

Figure \ref{fig:Fig8} shows the spatial electric field distributions on the object and image planes. Noise was artificially added to the fields on
the image plane that were calculated with COMSOL. The resultant noisy images are indicated by the red and green lines in the figure. The Tukey
(tapered cosine) window function was applied to the image plane only. The Fourier transforms of the images, with and without added noise are shown
in figure \ref{fig:Fig9}. The black line, which corresponds to $I_{N}(k_y)$ in Eq. \ref{eq:NOISE_PASSIVE}, shows how the added noise has clearly
affected the raw image spectrum beyond $\frac{k_y}{k_o} = 2$, where all of the object features are now completely buried under the noise and
indiscernible. However, the blue line, which corresponds to $I^{'}_{N}(k_y)$ in Eq. \ref{eq:NOISE_ACTIVE}, shows how these features can be
recovered with the convolved auxiliary.

We propose the following iterative process to apply the auxiliary source. We then use active compensation filter to reconstruct the image spectrum.
\begin{enumerate}
\item Select an arbitrary $k_c$  in the region where the noise has substantially degraded the spectrum. Choose a guess auxiliary amplitude by selecting $P_0$.
\item Convolve the object with $A(y)$ to obtain the total object.
\item Measure the electric fields on the image plane corresponding to the total object.
\item Re-scale $P_0$ for the selected $k_c$ if necessary, to make sure that adequate amplification is available in the image plane and noise is not visible in the Fourier spectrum.
\item Select another $k_c$ on the noise floor and ensure that there is sufficient overlap between the adjacent auxiliaries.
\item Repeat the processes in $1-5$ by superimposing those multiple auxiliaries until the transfer function of the imaging system is reasonably
accurate. We were restricted by the inaccuracy of the simulated passive transfer function $T_P(k_y)$ beyond $\frac{k_y}{k_o} = 4.7$ which prevented
us from going beyond.
\end{enumerate}
Note that in the above steps the selection of $P_o$ does not require prior knowledge about the object. The blue and green lines of
figure \ref{fig:Fig9} show the total images with and without added noise, respectively. They include four auxiliary sources with the center
frequencies $k_c = k_y/k_o = 2.8, \ 3.1, \ 3.6, \ 4.2$ and $P_o = 3000, \ 10000, \ 4 \times 10^5, \ 10^6$, respectively, with the
same $\sigma = 0.35$. The final $A(y)$ was then convolved with the object and the resulting total field distribution on the object plane is shown
by the blue line in figure \ref{fig:Fig8}.

\begin{figure}[htbp]
\centering
\includegraphics[width=\linewidth]{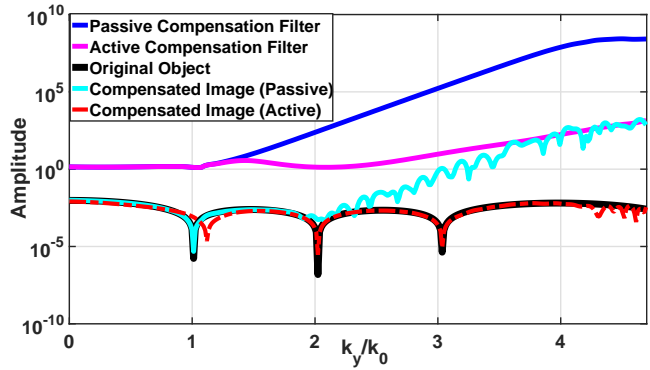}
\caption{Fourier spectra [$Vm^{-1}$] of the reconstructed images illustrating the difference between active and passive compensation. The passive compensation has significantly amplified the noise whereas the active one does not.}
\label{fig:Fig10}
\end{figure}

Active compensation filter, defined by Eq. \ref{eq:CFA} and illustrated in figure \ref{fig:Fig10}, is then multiplied in the spatial frequency
domain by the total image spectrum with the added noise. The resulting compensated spectrum is the red line of figure \ref{fig:Fig10}. The noise,
which was visible in the total image spectrum beyond $\frac{k_y}{k_o} =  4.5$, is also amplified in this reconstruction process.
However, in the regions where the auxiliary source is sufficiently strong, suppression of noise amplification is evident. The reconstructed spectrum
perfectly coincides with the original object shown by the black curve in figure \ref{fig:Fig10}. The light blue line corresponds to the passively
compensated image obtained by multiplying Eq. \ref{eq:CF} (i.e., dark blue line in figure \ref{fig:Fig10}) with the noise added raw image
(i.e., black line in figure \ref{fig:Fig9}). The advantage of the active compensation over its passive counterpart is therefore clearly evident
from the reconstructed Fourier spectrum. After the loss compensation process, the spectrum is truncated at $\frac{k_y}{k_o} = 4.7$ because the
simulated transfer function loses accuracy.

Figure \ref{fig:Fig7} shows that the passive transfer function starts to flatten beyond $\frac{k_y}{k_o} = 4.5$, even though the analytical transfer
function  monotonically decreases (see figure 3 in \cite{Chen:16}), inaccurate simulated transfer function $T_P(k_y)$ indicates that it is no longer
possible to perform the required compensation accurately (see Eq. \ref{eq:CFA}). More precisely, the imaging system requires more compensation than
the transfer function predicts. The reconstructed spectrum therefore starts to deviate from the original object when $\frac{k_y}{k_o} > 4.5$ as seen
in the red plot of figure \ref{fig:Fig10}, indicating inadequate compensation. This was one of the main reasons why we were unable to image beyond
$\frac{k_y}{k_o} = 5$.

Additionally, reconstructing the object features successfully requires a strong amplification. The auxiliary amplitude necessary to
produce this amplification is very high and it starts to generate substantial electric field oscillations towards the edges of the image
plane. Because the image plane is finite along the y-axis and the electric field is abruptly cut at a point where it is non-zero, a computational
error is introduced in the spatial Fourier transform. An artefact of this can be seen in figure \ref{fig:Fig10} where the red plot shows that the
feature at $\frac{k_y}{k_o} = 1$ is slightly shifted. The error is more prominent when the intensity of illumination is increased. Extending the
size of the image plane along the y-axis mitigates the error at the expense of computational or physical resources.

\begin{figure}[htbp]
\centering
\includegraphics[width=\linewidth]{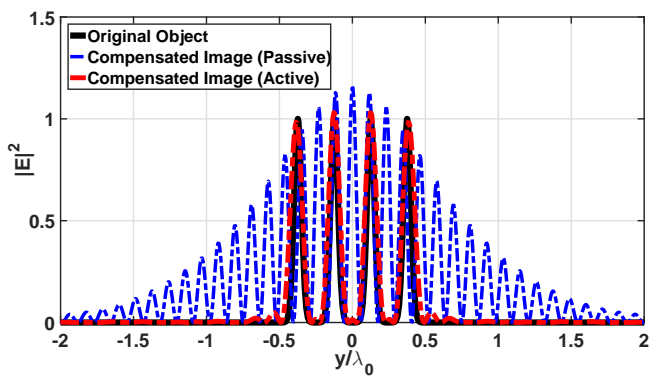}
\caption{Reconstructed images showing the difference between active and passive compensation schemes. Note that the passively compensated image has been scaled down by $10^7$.}
\label{fig:Fig11}
\end{figure}

Note that towards the tails of the amplification, or when $\frac{k_y}{k_o} > 4.5$, the amplification is not strong enough to overcome the noise.
Hence, there should be sufficient overlap between the two adjacent auxiliaries. The FWHM of $P(k_y)$, controlled by $\sigma$, can be selected
arbitrarily. In our simulations we were limited by the finite image plane. A very narrow $P(k_y)$ in the spatial frequency domain translates to
a wide field distribution in the spatial domain. This created additional field oscillations towards the edges of the image plane increasing
the errors in the Fourier transform calculations. Figure \ref{fig:Fig11} shows the amplitude squared of the reconstructed fields illustrating the
improvement of the active over passive compensation scheme.

\section{Discussion}

\begin{figure}[htbp]
\centering
\includegraphics[width=\linewidth]{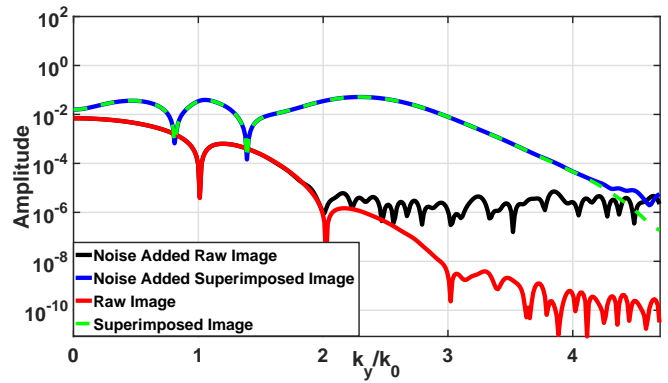}
\caption{Fourier spectra [$Vm^{-1}$] of the raw image superimposed with several object independent auxiliaries. Such superposition does not provide amplification and hence the feature at $\frac{k_y}{k_o} = 3$ is not recovered. Note that the black and red lines are the raw images with and without added noise, respectively, as also shown in figure \ref{fig:Fig9}.}
\label{fig:Fig12}
\end{figure}

The active compensation scheme works, because the convolved auxiliary source allows us to \emph{``selectively amplify''} spatial frequency features
of the object. This amplification cannot be achieved simply by the superposition of the object with an object independent auxiliary source. This is
illustrated in figure \ref{fig:Fig12} where we set $O(k_y) = 1 Vm^{-1}$ in the second term of Eq. \ref{eq:TOTALOBJ} and use the same $P(k_y)$
distributions in figure \ref{fig:Fig9}. The blue and green lines correspond to the images of the object superimposed with the object independent
auxiliaries with and without added noise, respectively. The buried object spectrum at $\frac{k_y}{k_o} = 3$ shown in figure \ref{fig:Fig9} has
gone undetected.

The convolution process to construct the auxiliary source that was described in this paper can be thought of as a form of structured light
illumination or wavefront engineering \cite{Zhao,Kildishev1232009,Pors2013,PhysRevB.84.205428,Yu333,Xu:16,smith2003limitations,Cao:17}. Along these
lines, for example, a plasmonic lens imaging system was discussed recently in \cite{Zhao}, where the authors described the fields on the image
plane by Eq. \ref{eq:TF} that contains an illumination function as a result of a phase shifting mask. Spatial filters based on hyperbolic
metamaterials \cite{Rizza:12,schurig2003spatial,wood2006directed} may be promising for the implementation of the proposed convolution.
For example, an object illuminated with a high intensity plane wave and projected on such spatial filters can physically implement the
convolved auxiliary source corresponding to the second term in Eq. \ref{eq:TOTALOBJ} and used in step 2 of the iterative reconstruction process.
Here, the spatial filter needs to be engineered to have a transfer function of the form similar to Eq. \ref{eq:PUMP}. The object
(i.e., such as an aperture based object illuminated by a plane wave) is to be placed on top of this additional metamaterial layer. The field
distribution at the exit of this layer would be the convolution of the object field distribution with the point spread function of the layer,
hence leading to the auxiliary source term in Eq. \ref{eq:TOTALOBJ} (i.e., second term). One way to engineer such a transfer function is with
the hyperbolic metamaterials which support high spatial frequency modes. In \cite{wood2006directed}, for example, the transmission coefficient for
the transverse magnetic waves in a hyperbolic medium was shown to have multiple peaks in the high spatial frequency region. The position of these
peaks can be tuned by changing the filling fraction or the thickness of the hyperbolic medium. If one has engineered a metamaterial with a transfer
function having one transmission peak $P_0$ around a certain spatial frequency (i.e., center spatial frequency $k_c$ in Eq. \ref{eq:PUMP}) and is
zero everywhere else, the iterative process where $P_0$ is re-scaled (i.e., to control amplification) is functionally equivalent to re-scaling the
amplitude of the plane wave $E_0$ illuminating the object. It is also worth mentioning here that this physically means actively adjusting the
coherent plasmon injection rate in the imaging system to compensate the losses as conceptualized in \cite{PI}. On the other hand, controlling the
center frequency will require multiple or tunable metamaterial structures where the transfer function can be tuned to show transmittance peaks at
different center frequencies. Therefore, it would be advantageous to have a broad transmittance in a physical implementation as long as the noise
amplification does not start to dominate. Another possible way to construct the necessary transfer function may be with the use of
metasurfaces \cite{Genevet:17}, which are ultrathin nanostructures fabricated at the interface of two media. The scattering properties of the
sub-wavelength resonant constituents of the metasurfaces can be engineered to control the polarization, amplitude, phase, and other properties of
light \cite{Kildishev1232009,Pors2013,PhysRevB.84.205428,Yu333,pfeiffer2013metamaterial,holloway2012overview}. This can allow one to engineer an
arbitrary field pattern from a given incident illumination \cite{pfeiffer2013metamaterial,holloway2012overview}.

In order to understand how the active compensation enhances the resolution limit of the NIFL we need a deeper understanding of the effect of the
noise on the ideal image spectrum. Eqs. \ref{eq:SD_Passive} and \ref{eq:SD_Active} tell us that the signal dependent noise is amplified
proportionally with the illumination. But since the active compensation seems to work so well, can we say that the noise is not amplified to the
same extent as the signal? Then, would it be possible to achieve the same results by simply increasing the intensity of the plane wave illuminating
the object? To address these questions we will consider below the \emph{``weak illumination,''} \emph{``structured illumination''}
and \emph{``strong illumination''} cases.

\begin{figure}[htbp]
\centering
\includegraphics[width=\linewidth]{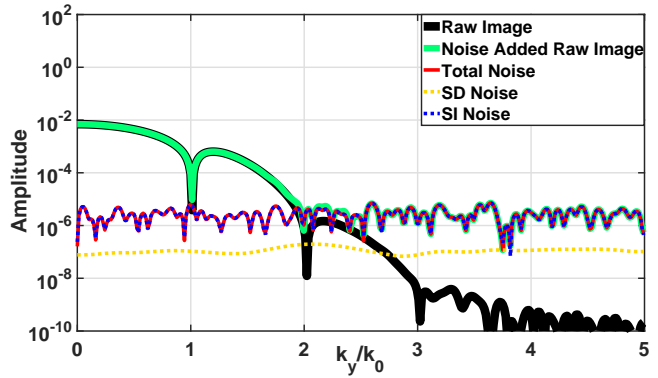}
\caption{Fourier spectra [$Vm^{-1}$] of the raw images with and without added noise illustrating the contribution of the SD and SI noise to the total distortion of the image under the weak illumination case.}
\label{fig:Fig21}
\end{figure}

\begin{figure}[htbp]
\centering
\includegraphics[width=\linewidth]{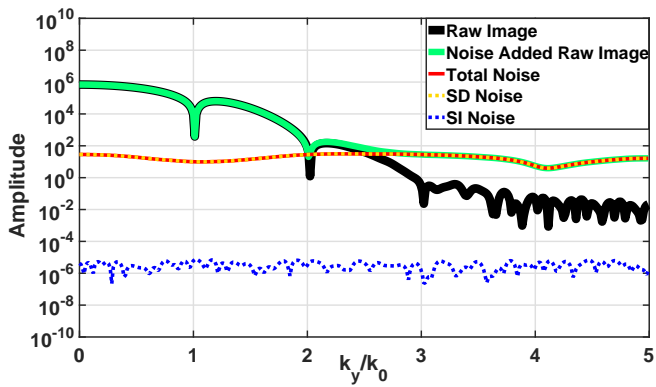}
\caption{Fourier spectra [$Vm^{-1}$] of the raw images with and without added noise illustrating the contribution of the SD and SI noise to the total distortion of the image under the strong illumination case. The SD noise is amplified approximately by a factor of $10^8$ throughout the spectrum and the contribution of the SI noise is very small.}
\label{fig:Fig22}
\end{figure}

\begin{figure}[htbp]
\centering
\includegraphics[width=\linewidth]{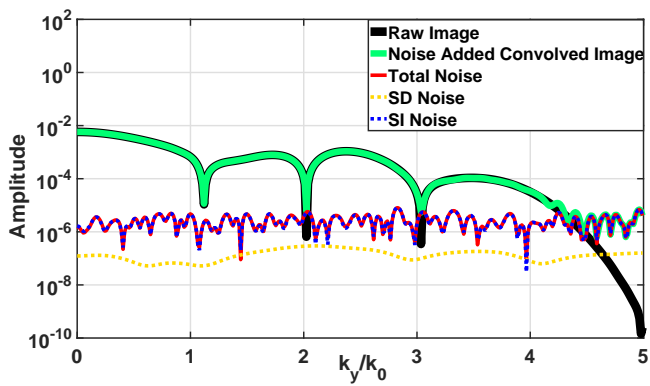}
\caption{Fourier spectra [$Vm^{-1}$] of the convolved images with and without added noise illustrating the contribution of the SD and SI noise to the total distortion of the image under the structured illumination case. The SD noise is approximately at the same level as the weak illumination except that the noise is redistributed.}
\label{fig:Fig23}
\end{figure}

We will take the Fourier transforms of Eqs. \ref{eq:NOISE_PASSIVE} and \ref{eq:NOISE_ACTIVE} and analyze how each noise term contributes to the total
distortion of the ideal image under the three illumination schemes. The linearity of the Fourier transform allows us to plot each term in the
equations separately and these are shown in figures \ref{fig:Fig21} - \ref{fig:Fig23} for the weak, strong, and structured illuminations,
respectively. In the strong illumination case we have used a plane wave whose electric field is $10^8$ times stronger than the weak illumination.
The green and black lines in figures \ref{fig:Fig21} - \ref{fig:Fig23} correspond to the images with and without added noise, respectively.
The Fourier transforms of the SI and SD noise are the blue and gold lines, respectively, which add up to the total noise shown by the red line.

To compare the performance of the imaging system under the three illumination schemes, we will see how closely the noise added images overlap with
the images with no added noise. We should note that the spatial distribution of  the SD noise  will be spread out over multiple Fourier
components \cite{4518410} and therefore the random nature of noise will not be visible in the spatial frequency domain. This can be seen in the
gold plots which are fairly smooth compared to the blue lines.


If we compare the SD noise spectra in figures \ref{fig:Fig21} and \ref{fig:Fig22} we immediately conclude that as we increase the intensity of the illumination, the SD noise is amplified throughout the spectrum. However, a slight improvement to the noisy spectrum over the weak illumination is visible within the region $2 < \frac{k_y}{k_o} < 3$. Additionally, if we analyze figure \ref{fig:Fig21}, where the gold line intersects the black line, we see that in the strong illumination case in figure \ref{fig:Fig22}, only the Fourier components until this intersection point are recovered. The intersection marks the spatial frequency at which the ideal image (i.e., raw image with no added noise) $I(k_y)$ matches the Fourier transform of the SD noise $N_{SD}(k_y)$. Beyond this point, we can say that the ideal image is completely buried under the SD noise alone. As we steadily increase the intensity of the illumination, $I(k_y)$ and $N_{SD}(k_y)$ increase by the same proportion and therefore, the value of $\frac{k_y}{k_o}$ where the two intersect does not change. We can therefore say that the improvement in the noisy spectrum in figure \ref{fig:Fig22} is due to the signal rising above the SI noise which does not change with the illumination intensity. Further increments in the strength of the illumination will not improve the noisy image spectrum.

On the other hand, if we study the SD noise spectrum in figure \ref{fig:Fig23}, we can see that it is approximately at the same average level as
figure \ref{fig:Fig21}. This is not surprising if we compare the green and red plots of figure \ref{fig:Fig8}. We can see that the spatial electric
field distribution of the noise added images $I^{'}_{N}(y)$ and $I_{N}(y)$ are comparable. The only difference is that $I^{'}_{N}(y)$ has high
spatial frequency features. Since the standard deviation of the SD noise is proportional to the amplitude of the image, according to
Eqs. \ref{eq:SD_Passive} and \ref{eq:SD_Active}, we can see why the SD noise is approximately the same in both the weak and structured
illumination schemes. Note that under the structured illumination, the noise added convolved image closely follows the ideal image
until $\frac{k_y}{k_o} = 4.5$. This can be pushed to even higher spatial frequencies if the transfer function characterizing the lens is
accurate. Also, note that the structured illumination has successfully suppressed the computational errors in the Fourier transform which
are visible in the black line in figure \ref{fig:Fig21} beyond $\frac{k_y}{k_o} = 3$. These errors are amplified by a factor of $10^8$ times
in figure \ref{fig:Fig22}.

From the above discussion we conclude that by using structured illumination the SD noise is not amplified but redistributed when compared with the
strong illumination. Therefore, it is possible to raise the high spatial frequency features of the object above the noise.  This is the primary
reason why structured illumination can accurately resolve the image while strong illumination fails.

The technique is generally applicable to any arbitrary object with the use of any plasmonic or metamaterial lens provided that accurate transfer
function for the imaging system is available. A selective amplification process is used to recover specific object features by controlling $P_0$ near
and beyond where the noise floor is reached in the Fourier spectrum of the raw image. Therefore, no prior knowledge of the object is required.
However, a necessary criterion is a sufficiently accurate transfer function in the region where the auxiliary is applied to correctly estimate the
required amount of amplification. It should be noted that different objects may require different auxiliaries, since the spatial frequency at which
the noise floor is reached may vary for different objects. Therefore, it would be instrumental to have a tunability mechanism for the versatility
of the imaging system. Even though a single narrowband auxiliary would be still sufficient to enhance the resolution of the raw image, further
enhancement in the resolution would demand either superimposing multiple narrowband auxiliaries or a single sufficiently broadband auxiliary within
the range of accurate transfer function. Narrowband auxiliaries require larger image plane and more post-processing while a single broadband
auxiliary requires less post-processing and smaller image plane at the expense of possibly higher noise amplification. Similarly, unnecessarily
large amplitude of a narrowband auxiliary may excessively amplify the noise. Another likely limitation arises from increasingly large power loss
in the deep subwavelength regime, which requires increasingly high amount of amplification to reconstruct extremely fine details of the object.
This does not only reduce the efficiency but might also introduce undesired non-linear and thermal effects in the optical materials, hence,
limiting the resolution of the imaging system.

\section{Conclusion}
In summary, we proposed an active implementation of the recently introduced plasmon injection scheme \cite{PI} to significantly improve the
resolution of Pendry’s non-ideal negative index flat lens beyond diffraction limit in the presence of realistic material losses and SD noise. Simply
by increasing the illumination intensity, it is not generally possible to efficiently reconstruct the image due to the noise amplification. However,
in the proposed active implementation one can counter the adverse noise amplification effect by using a convolved auxiliary source which allows for
a selective amplification of the high spatial frequency features deep within the sub-wavelength regime. We have shown that this approach can be used
to control the noise amplification while at the same time recover features buried within the noise, thus enabling ultra-high resolution imaging far
 beyond the previous passive implementations of the plasmon injection scheme \cite{Wyatt,zhang2016enhancing}. The convolution process to construct
 the auxiliary source in the proposed active scheme may be realized physically by different methods,
 metasurfaces \cite{Genevet:17,Kildishev1232009,Pors2013,PhysRevB.84.205428,Yu333,Xu:16,pfeiffer2013metamaterial,holloway2012overview} and
 hyperbolic metamaterials \cite{Rizza:12,schurig2003spatial,wood2006directed,zhang2015hyperbolic} being the primary candidates. A more detailed
 analysis on the design of such structures to implement the convolved auxiliary source will be the focus of our future research. Finally, we should
 note that we purposefully focused on imperfect negative index flat lens here that poses a highly stringent and conservative problem. However, in
 the shorter term the proposed method can be relatively easily applied to experimentally available plasmonic
 superlenses \cite{Guo2014,Liu_Hong,Zhao,Fang534,taubner2006near,zhang2008superlenses} and
 hyperlenses \cite{liu2007far,rho2010spherical,lu2012hyperlenses,sun2015experimental,lee2007development}. Our findings also raises the hopes for
 reviving Pendry’s early vision of perfect lens \cite{Pendry} by decoupling the loss and isotropy issues toward a practical
 realization \cite{soukoulis2010optical,soukoulis2011past,guney2009connected,guney2010intra,rudolph2012broadband,yang2016experimental}.

\section{Funding Information}
Office of Naval Research (award N00014-15-1-2684).

\section{Acknowledgement}

We thank Jeremy Bos at Michigan Technological University for fruitful discussion on noise characterization.

\end{document}